\documentclass{iaus}
\usepackage{graphicx}

\title[Magnetic Towers and Binary-Formed Disks] 
{Magnetic Towers and Binary-Formed Disks: New Results for PN Evolution}

\author[Huarte-Espinosa \&, Frank]   
{Mart\'{\i}n Huarte-Espinosa \and Adam Frank}

\affiliation{Department of Physics and Astronomy, University of Rochester, \\
600 Wilson Boulevard, Rochester, NY, 14627-0171 \\ 
emails: {\tt martinhe; afrank; @pas.rochester.edu}}

\pubyear{2011}
\volume{283}  
\pagerange{119--126}
\jname{Planetary Nebulae: an Eye to the Future}
\editors{A.C. Editor, B.D. Editor \& C.E. Editor, eds.}
\begin{document}

\maketitle

\begin{abstract}
We present new results of 3-D AMR MHD simulations focusing on two
distinct aspects of PPN evolution. We first report new simulations
of collimated outflows driven entirely by magnetic fields. These
Poynting flux dominated ``magnetic towers'' hold promise for explaining
key properties of PPN flows. Our simulations address magnetic tower
evolution and stability. We also present results of
a campaign of simulations to explore the development of accretion
disks formed via wind capture. Our result focus on the limits of
disk formation and the range of disk properties.
\end{abstract}

\firstsection 
\section{Introduction}

Jets are observed in proto-Planetary Nebuale (PPN), Young Stellar Objects,
radio galaxies and other astrophysical objects.
Models suggest that jets are launched and collimated by accretion,
rotation and magnetic mechanisms (\cite[Pudritz et al. 2007]{pudritz}).
One paradigm to explain the formation of jets in PPN is that of a
binary system in which material from an AGB star is accreted onto
an intermediate-mass companion (\cite[Soker \& Rappaport 2000]{soker};~
\cite[Nordhaus, Blackman \& Frank 2007]{jason}).  Jets or collimated
outflow have recently been observed in the close binaries NGC~6326
and NGC~6778 (\cite[Miszalski et al. 2011]{misz}).  Previous numerical studies
of wind capture in binaries have found enhancements over Bondi-Hoyle
accretion rates onto the secondary (\cite[Mastrodemos \& Morris
1998]{mm}; \cite[Podsiadlowski \& Mohamed 2007]{pod}; \cite[De Val-Borro et al. 2009]{borr}).
These results, however, must be tested using high resolution simulations
because the implications for the maximum outflow power are dramatic;
the answer could rule in, or out, the secondary as the engine
powering jets in PPN.

In the case of magnetized jets, plots of specific angular momentum
vs. jet poloidal speed are distinct for different MHD engines
(\cite[Ferreira et al. 2006]{ferr}).  This fact can help constrain the physics of
the jet engine.  The importance of the magnetic fields relative to
the flows' kinetic energy, divides jets into (i) Poynting flux
dominated (PFD; \cite[Shibata \& Uchida 1986]{shibata}), in which
magnetic fields dominate the jet structure, (ii) magnetocentrifugal
(\cite[Blandford \& Payne 1982]{bland}), in which magnetic fields
only dominate out to the Alfv\'en radius.  The observable differences
between PFD and magnetocentrifugal jets are unclear, as are the
effects of cooling and rotation on PFD jets.  Recently, cooling
magnetized jets have been formed in laboratory experiments
(\cite[Lebedev et~al. 2005]{leb}). Such studies, along with high
resolution 3-D MHD numerical simulations, have proven to play a key
role in the understanding of the physics of jet launch and stability.

\section{Magnetic towers}
\subsection{Model}
We use the Adaptive Mesh Refinement (AMR) code AstroBEAR2.0 
(\cite[Cunningham et~al. 2009]{bear};~
\cite[Carroll-Nellenback et~al. 2011]{bear2}) 
to solve the equations of radiative-MHD in 3D.
The grid represents 160$\times$160$\times$400\,AU
divided into 64$\times$64$\times$80 cells plus 2 AMR levels. Initially,
the molecular gas is static and has an ideal gas equation of state
($\gamma=\,$5$/$3), a number density of 100\,cm\,s$^{-1}$ and 
a temperature of 10000\,K.
The magnetic field is helical, centrally localized and given 
by the vector potential (in cylindrical coordinates) 
${\bf A}(r,z) =  [(r/4)(cos(2r) + 1)( cos(2z)
+ 1 )] \hat{\phi} + [5\,(cos(2r) + 1)( cos(2z) + 1 )] \hat{k}$,
for $r,z <$ 30\,AU, and ${\bf A}(r,z) =\,$0 elsewhere. The magnetic
pressure is higher than the thermal one inside the magnetized
region. 

Source terms are used to continually inject magnetic or kinetic energy 
at cells $r,z<\,$30\,AU.
We carry out 4 simulations: an adiabatic, a cooling (\cite[Dalgarno
\& McCray 1972]{dm}) and a rotating (Keplerian) PDF jet,
as well as a hydrodynamical jet which has the same time average
propagation speed and energy flux than the adiabatic PFD jet.

\begin{figure}[hb]
\begin{center}
\includegraphics[width=0.7\textwidth]{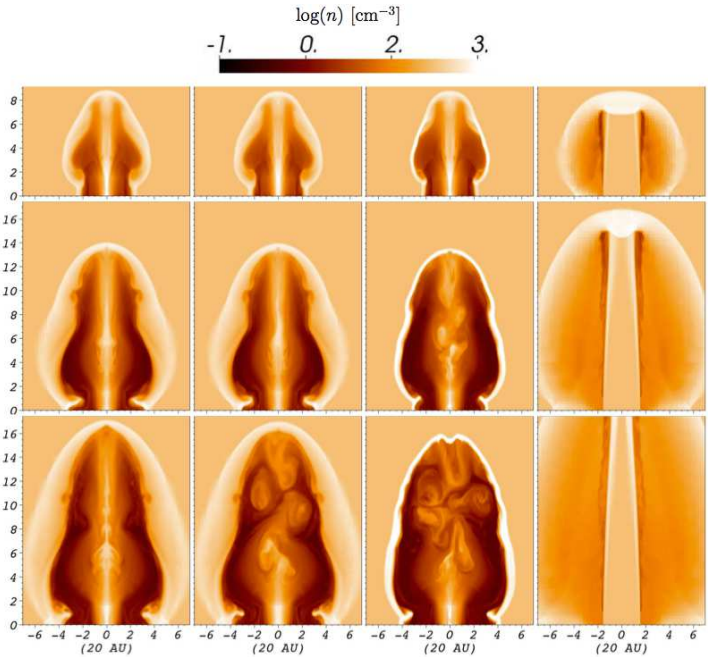} \\
\end{center}
\vspace{-0pt}
\caption{Logarithmic false color number density (cm$^{-3}$) maps
of the gas at the midplane of the cubic computational domain. 
Adiabatic (1st column), rotating
(2nd column) and cooling (3rd column) PDF jets. The hydrodynamic jet 
is in the 4th column. From top to bottom the time is 42, 84 and 118\,yr.}
\end{figure}

\subsection{Results}
Magnetic pressure pushes field lines and plasma up, forming magnetic
cavities with low density. The adiabatic case is the most stable.
PFD jets decelerate relative to the hydrodynamic one; magnetic
energy pressure produces axial but also radial expansion. PFD jet
cores are thin and unstable, whereas the hydrodynamic jet beam is
thicker, smoother and stable. The PFD jets are sub-Alfv\'enic. Their
cores are confined by magnetic hoop stress, while their surrounding
cavities are collimated by external thermal pressure. PDF jets carry
high axial currents which return along their outer contact
discontinuity. Pinch, $m=\,$0, kink, $m=\,$1, and $m=\,$2 current-driven
perturbations are evident in the PDF jets.  The perturbations are
amplified by cooling, firstly, and base rotation, secondly.  This
happens because both shocks and thermal pressure support are weakened
by cooling, and the total pressure balance at the jets' base is
affected by rotation. See \cite[Huarte-Espinosa et al. 2011]{tower} (in prep)
for details.

\section{Binary-formed disks}
\subsection{Model}
We use Astrobear2.0 (\cite[Cunningham et~al. 2009]{bear};~
\cite[Carroll-Nellenback et~al. 2011]{bear2}) to solve the equations
of hydrodynamics in 3D.  The grid represents 100\,AU$^3$ divided
into 32$^3$ cells. We also use 5 AMR levels which allow us to have high
resolution, of order 0.4\,AU, around the model stars. These, are implemented with
gravity particles, separated by 25\,AU and follow circular orbits.
The primary simulates an AGB star with a mass of 1.5\,M$_{\odot}$, and a
spherical constant wind of 10\,km\,s$^{-1}$ and $\dot{M}
\sim\,$10$^{-8}$\,$M_{\odot}$\,yr$^{-1}$.  The secondary star
simulates a main sequence star or a white dwarf with 1\,M$_{\odot}$.

\begin{figure}[ht]
\begin{center}
\includegraphics[width=0.7\textwidth]{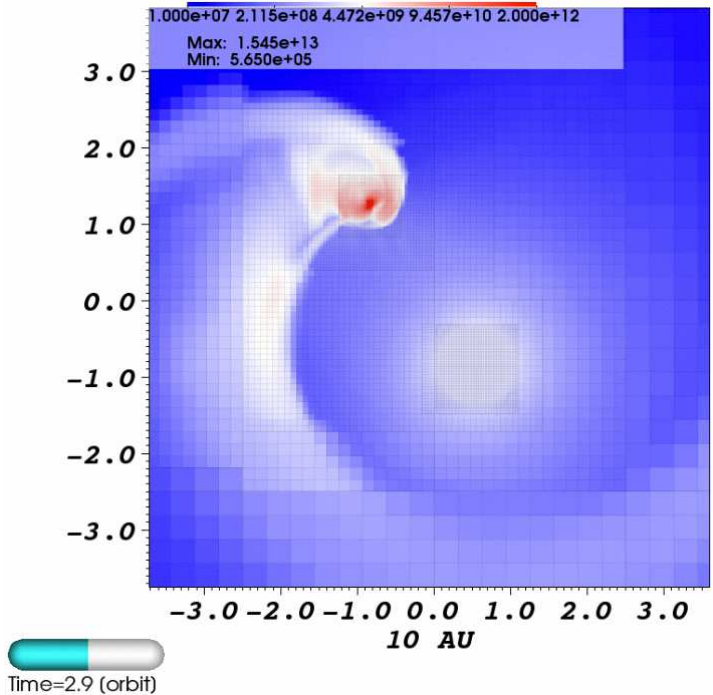} \\
\vspace{0pt}
\caption{Logarithmic number density (cm$^{-3}$) map of the gas in
the orbital plane. The primary and secondary stars are located at
(0.6,-0.9) and (-0.9,1.2), respectively. The blue and white components
are the AGB wind, the spiral structure of which is caused by the orbital
motion. The red gas component is the accretion disk that is formed
around the secondary via wind capture.
}
\end{center}
\vspace{-10pt}
\end{figure}

\vskip0.5cm
Initially, we implement an isothermal ($\gamma=1.001$) gas in the grid which is unaffected by the
gravitational field of either of the stars, but they pull each
other around the center of gravity which is located
at the origin (Figure~2).  The system orbits twice and the grid is filled with
material from the primary's AGB wind.  The orbital motion 
causes a helical-like distribution on the wind.  We then
``turn-on'' the effect of the secondary's gravity on the gas and
allow the system to orbit ten times under these conditions.  

\subsection{Results}
As the stars orbit each
other, the dense gas from the AGB wind flowing near 
the secondary is captured by the star's gravitational
field. This process is essentially a form of Bondi-Hoyle flow and the flow pattern is localized near the secondary.
A spiral shock pattern does however propagate through the AGB wind.  

Within one orbital timescale we see that an
accretion disk forms around the secondary.  We note that previous
3D simulations of accretion disk formation via wind capture in binaries
have explored star separations smaller than our model with 25\,AU 
(e.g. \cite[Mastrodemos \& Morris 1998]{mm}). Thus our results show that disks can form
even out to large separations.

Preliminary analysis finds the density and temperature of the disk progressively increasing in time.  We
also find that disk material close to the secondary follows Keplerian orbits confirming it is bound to the star.
See \cite[Huarte-Espinosa et al. 2012]{tower} (in prep) for details.

\section{Conclusions and Summary}
PFD jet beams are lighter, slower and less stable than kinetic-energy
dominated ones.  We predict characteristic emission distributions
for each of these.  Current-driven perturbations in PFD jets are
amplified by cooling, firstly, and base rotation, secondly:
shocks and thermal pressure support are weakened by cooling.  Total
pressure balance at the jets' base is affected by rotation.  Our
simulations are in good agreement with the models and experiments of
\cite[Shibata \& Uchida 1986]{shibata} and \cite[Lebedev et~al. 2005]{leb},
respectively.

Regarding the problem of the accretion disk formed by wind capture,
we see the formation of accretion disks at large binary separations 
of order 25\,AU. The density
and temperature of the disk progressively increases in time.

\begin{discussion}

\end{discussion}

\end{document}